\documentclass[fleqn,usenatbib]{mnras}

\usepackage{newtxtext,newtxmath}
\usepackage[T1]{fontenc}

\DeclareRobustCommand{\VAN}[3]{#2}
\let\VANthebibliography\thebibliography
\def\thebibliography{\DeclareRobustCommand{\VAN}[3]{##3}\VANthebibliography}
\newcommand\asloth{\textsc{a-sloth}}
\newcommand{\Msun}{\,\ensuremath{\mathrm{M}_\odot}}

\usepackage{graphicx}	% Including figure files
\usepackage{amsmath}	% Advanced maths commands
\title[Local value of the baryonic streaming velocity]{First estimate of the local value of the baryonic streaming velocity}

\author[Uysal \& Hartwig]{
Betül Uysal,$^{1, 2}$ \thanks{E-mail: betul.u@outlook.fr}
Tilman Hartwig,$^{2, 3, 4}$
\\
$^{1}$Department of Physics, University of Orléans, 1 rue de Chartres, 45100 Orléans, France\\
$^{2}$Department of Physics, School of Science, The University of Tokyo, Bunkyo, Tokyo, 113-0033, Japan \\
$^{3}$Institute for Physics of Intelligence, School of Science, The University of Tokyo, Bunkyo, Tokyo, 113-0033, Japan \\
$^{4}$Kavli Institute for the Physics and Mathematics of the Universe (WPI), The University of Tokyo Institutes for Advanced Study,\\
The University of Tokyo, Kashiwa, Chiba, 277-8583, Japan\\
}

\date{Accepted XXX. Received YYY; in original form ZZZ}

\pubyear{2022}

\begin{document}
\label{firstpage}
\pagerange{\pageref{firstpage}--\pageref{lastpage}}
\maketitle

\begin{abstract}
Several studies have shown the influence of the relative streaming velocity (SV) between baryons and dark matter on the formation of structures. For the first time, we constrain the local value of the SV in which the Milky Way was formed.
We use the semi-analytical model \asloth\ to simulate the formation of Milky Way-like galaxies. The high resolution in mass and time of the dark matter merger trees from the Caterpillar simulation enables us to accurately model star formation in the smallest progenitor halos at high redshift. The efficient semi-analytical nature of \asloth\ allows us to run many simulations with various values of the local SV.
Our investigation on the influence of the SV shows that it delays star formation at high redshift. However, at redshift z=0, the SV has no effect on the total stellar mass in the Milky Way nor its Satellites. We find that extremely metal-poor and ultra metal-poor stars are affected by the SV, and can hence be used to constrain its local value. The local optimal value of the SV is $v_\mathrm{SV} =1.75^{+0.13} _{-0.28}\,\sigma_\mathrm{SV}$, which is based on four independent observables. We further find that the SV decreases the number of luminous Milky Way satellites, but this decrease is not enough to solve the missing satellite problem.

\end{abstract}


%%%%%%%%%%%%%%%%%%%%%%%%%%%%%%%%%%%%%%%%%%%%%%%%%%

%%%%%%%%%%%%%%%%% BODY OF PAPER %%%%%%%%%%%%%%%%%%

\section{Introduction}
Dark matter (DM) halos are fundamental structures to understand the formation of galaxies \citep{barkana_beginning_2001}. To explain the formation of our own Galaxy, it is therefore crucial to study the interaction of DM and baryonic physics.

One phenomena that affects the behaviour of baryons inside DM halos is the baryonic streaming velocity (SV). Before the recombination epoch, photons and baryons are coupled by radiation pressure, which prevents the baryons from falling into the DM potential wells. Once the decoupling occurred around redshift $z=1100$, baryons could stream freely, but were affected by the already present structure of DM. The resulting SV between baryons and DM constitute a second-order effect in the structure equations \citep{tseliakhovich10,fialkov_supersonic_2014} and was hence often neglected previously.

Several studies have shown the importance of baryonic streaming velocities on the formation of structures in the high-redshift Universe. \citet{schauer_influence_2021} and \citet{kulkarni21} use numerical simulations to find the dependence of the critical mass above which a halo can form stars on the SV and show that a higher value of the SV requires a higher halo mass to retain gas and allow star formation. The SV can cause a spatial offset between density peaks of gas and DM \citep{naoz14}. In such offset gas density peaks, gas-rich objects can form, which are coined Supersonically Induced Gas Objects (SIGOs) and proposed as potential progenitors of globular clusters \citep{chiou19,schauer21,chiou21,Nakazato_2022,lake22,williams22}. The fraction of baryons in DM halos can be reduced by the SV below the cosmic mean baryon fraction in low-mass halos at high redshift \citep{greif11,OLeary12,naoz_simulations_2013,richardson13,asaba16,park20}. Several authors investigate the possibility that SV might have facilitated the formation of the first massive black hole seeds \citep{tanaka13,latif14,schauer17,hirano17}. The general influence of the SV on the formation of the first stars and galaxies has also been discussed in various works \citep{stacy11,ahn18,schauer19,druschke20}.
While these previous studies mostly focused on the high-redshift Universe, we want to extend these approaches and demonstrate the effect of the SV on local observables in the Milky Way (MW). Eventually, we also use these observables to constrain the local value of the SV.

The SV is defined as the relative bulk motion between baryons and cold DM
\begin{equation}
    \mathbfit{v}_\mathrm{SV} (\mathbfit{x}) = \mathbfit{v}_\mathrm{b} (\mathbfit{x}) - \mathbfit{v}_\mathrm{cdm} (\mathbfit{x}).
\end{equation}
The distribution of its magnitude, $v_\mathrm{SV}$, with position $\mathbfit{x}$ follows a Maxwell-Boltzmann distribution with variance $\sigma ^2 _\mathrm{SV}$ \citep{tseliakhovich10}. The value of this variance decays with redshift, following the relation
\begin{equation}
    \sigma _\mathrm{SV} (z) = 30\,\mathrm{km\,s}^{-1} \left( \frac{1+z}{1100} \right).
\end{equation}
Also the local values of the SV decay with redshift. Hence, once we know the value of the SV as multiple of the variance at one point at a given redshift, we know the entire evolution with redshift. Therefore, it is convenient to express the value of the baryonic streaming velocity at a certain point in the Universe as redshift-independent multiple of the local, redshift-dependent variance of the SV. For example, the most likely value for the SV is $v_\mathrm{SV} = 0.8 \sigma _\mathrm{SV}$ (maximum of the Maxwell-Boltzmann distribution).

In this study, we want to find the local value of the SV for the cosmic volume out of which the MW has formed. The coherence length of the SV is $\sim 3$ comoving Mpc, i.e., the baryonic streaming velocity can be considered constant on scales smaller than this coherence length \citep{tseliakhovich10,fialkov_supersonic_2014,asaba17}. The MW has formed out of an effective volume with side length $2-6$ comoving Mpc \citep{griffen_caterpillar_2016}, which is sufficiently close to the coherence length of the SV. Therefore, we can consider that the MW formed out of a region of the Universe that has one coherent streaming velocity. Since the effective volume, out of which the MW has formed, might be slightly larger than the coherence length of the SV, we will consider this in the discussion section, when we determine the precision with which we can determine the local value of the SV.

Studying the effects of the SV on the present-day MW and determining its local value has several important implications. As we are going to show, some local observables depend on the SV. Therefore, to maximize the information gain from such available observables, we have to constrain the local value of the SV, which has shaped the formation of the MW. Moreover, we show that the SV can mitigate the missing satellite problem. However, the exact suppression of satellite galaxies due to baryonic streaming depends on the exact value of the SV.

\section{Methodology}
In order to simulate the formation of galaxies, different methods are used, such as hydrodynamical simulations or semi-analytical models. We need to simulate the formation of the MW from very high redshift to the present day with sufficient mass resolution to resolve halos of mass $\sim 10^6\Msun$ at high redshift. In addition, we want to repeat this simulation many times for different values of the local SV. Such a suite of simulations is only feasible with semi-analytical models that allow large parameter explorations thanks to their modest runtime. We use the semi-analytical model \asloth\ \citep{Hartwig22,magg22} for this research, which is ideally suited to connect physics from the high-redshift Universe to observables in the MW.

\subsection{A-SLOTH}\label{sec:asloth}
\asloth\ is a public semi-analytical model\footnote{\url{https://gitlab.com/thartwig/asloth}}. The aim of this model is to simulate the formation of stars and galaxies and to compare them to observables. More specifically, it is optimized to connect the formation of the first stars and galaxies to local observables. It models the evolution of the Universe from high redshift to the present day. It is based on DM merger trees, which are obtained by the Extended Press-Schechter theory, or by using merger trees from DM simulations, such as the \textit{Caterpillar} trees that we will discuss in section~\ref{Caterpillar}. Their high mass resolution allows us to follow the formation of the first minihalos at high redshift, which are often unresolved in hydrodynamical simulations of comparable volume due to limited computational resources.

Once a DM halo crosses the critical mass for star formation, $M_\mathrm{crit}$, we start the baryonic cycling for this branch. Specifically, we first assign hot gas to this halo, according to the cosmic mean baryon fraction. Then, this hot gas cools to become cold gas, which can turn into stars once the amount of cold gas is higher than the local Jeans mass. Depending on the gas metallicity and chemical composition, we form Pop~III or Pop~II stars \citep{chiaki17}. Stars are then sampled randomly following pre-defined Initial Mass Functions (IMFs). Depending on their lifetime and mass, a star emits ionising photons and might eject elements in a supernova (SN) explosion at the end of its lifetime. The SN explosion energy drives galactic outflows and the ejected metals are mixed with the ISM and IGM \citep{tarumi20}. Thanks to the sampling of individual stars, we can trace the lifetimes and metallicities of all stars and sample accurately the metallicity distribution function (MDF) at redshift $z=0$.

\asloth\ has various free input parameters. The relevant input parameters are the SV ($v_\mathrm{SV}$) in units of $\sigma _\mathrm{SV}$, the maximum mass of the Pop~III IMF ($M_\mathrm{max}$), the Pop~II and Pop~III star formation efficiencies ($\eta _{III}$, $\eta _{II}$), the slope and normalisation of the supernova feedback efficiency ($\alpha _\mathrm{out}$, $M_\mathrm{out,norm}$), and the escape fraction of ionizing radiation from Pop~II stars ($f_\mathrm{esc,II}$).

Among these input parameters, the value of the SV is special. While the other input parameters summarize or quantify some subgrid physics that is not included in the baryonic modelling of \asloth, the value of the SV is an external cosmic parameter, which is fixed for a given physical volume, but usually unknown. Our task in this paper is to find this local value of the SV, which best reproduces observations, and to show that this local value is robust against the choice of the other input parameters.

\subsubsection{Observables}
The input parameters of \asloth\ are calibrated based on six observables.  Four of these observables are specific for the MW and therefore should also reflect any influence of the local value of the SV. The other two observables, which were used to calibrate \asloth, are the optical depth to Thomson scattering and the cosmic star formation rate density at $z \gtrsim 5$. These two observables require large cosmological volumes to be reproduced and are therefore not representative for our rather small MW-like simulations and do not reflect the local volume of the SV. Hence, in this present paper, we only focus on the remaining four observables, which we summarize in the following.
\begin{itemize}
    \item The stellar mass of the MW is $M_\mathrm{MW} = (5.43 \pm 0.57) \times 10^{10}\Msun$ \citep{McMillan17}.
    \item The cumulative stellar mass function (CSMF) of the MW satellite galaxies is sensitive to the star formation histories of MW satellites, which have rather old stellar populations. The completeness limit for current Satellite surveys is a stellar mass of $\gtrsim 5 \times 10^5\Msun$. In this range, we compare the distribution of observed satellite stellar masses to the ones simulated with \asloth\ and calculate the p-value with the two-sample Kolmogorov-Smirnov test. We further discuss the stellar mass function of MW Satellites in  Sec.~\ref{sec:MissingSat}.
    \item The fraction of extremely metal-poor (EMP, [Fe/H]$<$-3) stars in the MW halo compared to all stars in the MW quantifies the relative amount of early star formation at high redshift and low metallicity. We use an observed value of log$_{10}$(EMP/All) $=-4.7 \pm 0.5$ \citep{bullock05,bell08}.
    \item The ratio of ultra metal-poor (UMP, [Fe/H]$<$-4) to EMP stars quantifies the slope of the low-metallicity tail of the MDF. Observations find a value of log$_{10}$(EMP/UMP) $=(1.4-2) \pm 0.3$ \citep{youakim20,chiti21,bonifacio21}. This ratio is more robust than for example UMP/All, because EMP/UMP is independent of the rather uncertain mass of the stellar halo \citep{chen22}.

\end{itemize}
To quantify how well \asloth\ reproduces the MW, we first calculate individual p-values for the four observables, $p_i$, which we then multiply to obtain a final goodness-of-fit parameter
\begin{equation}
    p = p_1 \times p_2 \times p_3 \times p_4.
\end{equation}
We use the p-value from the Kolmogorov-Smirnov test for the CSMF, which quantifies the agreement between model and observations, and we use the p-values from the chi-square distribution for the other three observables for which we have mean values and uncertainties. For a more detailed discussion of these observables, see \citet{Hartwig22}.

The individual p-values are between 0$\le p_i \le$1, and are motivated by hypothesis testing: conventionally, one defines a threshold (e.g. $0.05$) below which one rejects the null-hypothesis that the model and observed data are drawn from the same underlying distribution. Multiplying these individual p-values has no rigorous mathematical definition since we can not guarantee that the four individual p-values are statistically independent. Therefore, we call $p$ goodness-of-fit parameter, which has the desired properties: maximising it will improve the agreement between model and observations. In a future study, we will express all observables as likelihoods, which have a more robust definition.

\subsubsection{Influence of SV}
The main influence of the SV is the critical halo mass. To form stars, gas within a halo has to cool by losing energy via photon emission. There are two dominant cooling channels at high redshift, the atomic H cooling and the molecular H$_2$ cooling \citep{glover13}. The critical halo masses for the atomic and molecular cooling channel differ, as their cooling efficiency is different \citep{barkana_beginning_2001}.
We can then define the critical halo mass, M$_{\textrm{crit}}$, for when star formation can occur as the minimum mass of the molecular cooling halo mass \citep{schauer_influence_2021}
\begin{equation}
    \log_{10}(M_1/M_{\odot}) = 6.017(1.0+0.166\sqrt{J_\mathrm{LW}/J_{21}})+0.416(v_{SV}/\sigma _\mathrm{SV}),
\end{equation}
and the atomic cooling halo mass \citep{hummel12}
\begin{equation}
M_2/M_{\odot} = 10^{7.5}\left(\frac{1+z}{10}\right)^{-1.5}.
\end{equation}
The volume of a MW-like galaxy is not sufficiently big to simulate the self-consistent build up of an external Lyman-Werner (LW) background $J_\mathrm{LW}$. Therefore, we apply the LW background based on \citet{greif06} in the form $J_\mathrm{LW}=10^{2-z/5}J_{21}$, where $J_{21} = 10^{-21} \mathrm{erg\,s}^{-1} \mathrm{cm}^{-2} \mathrm{Hz}^{-1} \mathrm{sr}^{-1}$.
We finally obtain for the critical mass
\begin{equation}
M_{\textrm{crit}}= \min(M_1,M_2).
\end{equation}

\begin{figure}
    \centering
    \includegraphics[width=\columnwidth]{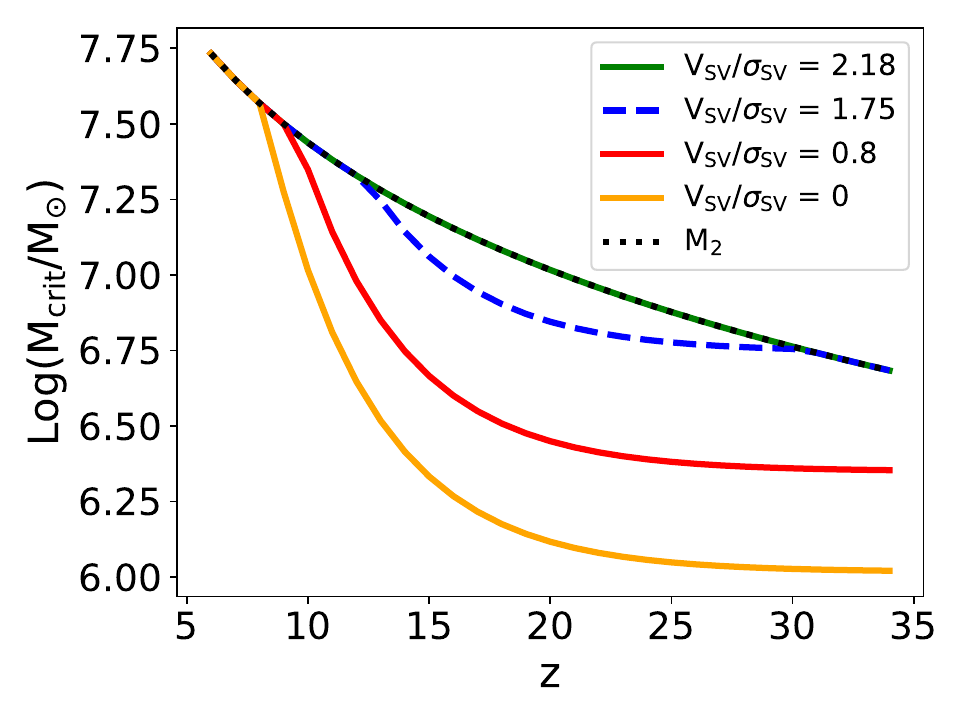}
    \caption{Redshift evolution of the critical halo mass for different streaming velocity. We see that larger SV result in higher critical masses. At lower redshift, the critical halo mass behaves independently from the SV. The black dotted line shows the atomic cooling mass, $M_2$. The value of $v_\mathrm{SV}/\sigma _\mathrm{SV} = 1.75$ (blue dashed line) is the optimal local value, as we show below.}
    \label{fig:halo_mass}
\end{figure}

We see in the Fig.~\ref{fig:halo_mass} the redshift evolution of the critical halo mass for different SV. For instance, the critical mass of $v_\mathrm{SV}=2.18\sigma_\mathrm{SV}$ is higher than the one for $v_\mathrm{SV}=0.8\sigma_\mathrm{SV}$.
We can also see that $M_1>M_2$ for $v_\mathrm{SV}/\sigma _\mathrm{SV} > 2.18$ at all redshifts and therefore $M_\mathrm{crit} = M_2$. In other words, $v_\mathrm{SV}/\sigma _\mathrm{SV} \sim 2.18$ is the maximum value to which the SV has an effect on minihalos, which are cooled by molecular hydrogen. Above this threshold, star formation is delayed until halos reach the atomic cooling mass, $M_2$, and can cool via this more efficient cooling channel. Therefore, we use $v_\mathrm{SV}/\sigma _\mathrm{SV} \sim 2.18$ as the maximum explored value because the results of our simulation would not change if the SV would be higher than this threshold.

The SV only affects baryonic physics at the highest redshift. Therefore, it is important to model star formation at high redshift very accurately in order to model any influence of the SV on present-day observables.

\subsection{Caterpillar Merger Trees}\label{Caterpillar}
The exact merger and formation history of the MW is still uncertain. To obtain a realistic merger tree for a MW-like galaxy, we therefore use the data from the \textit{Caterpillar} project, which is a simulated set of 30 DM halos \citep{griffen_caterpillar_2016}. These merger trees provide high spatial ($\sim 10^4\Msun$ per DM particle) and temporal (5\,Myr per snapshot) resolution.

Each merger tree contains DM halos, which represent the history of a MW-like galaxy. This means that the simulated galaxy has a similar DM mass to the MW, there was no major merger since $z=0.05$, and no second massive halo within $\sim 2.8$\,Mpc. In order to identify the halos, they used a modified version of the halo-finder algorithm  \textsc{rockstar} \citep{BehrooziROCKSTAR}, and to construct the merger trees, they used \textsc{consistent-trees} \citep{behrooziConstistent}.

The 30 merger trees have sufficient mass resolution to resolve minihalos at the highest redshift, they start at redshift $\sim 30$, and they contain Satellite galaxies within the virial radius of the MW. This selection of MW-like merger trees is therefore ideally suited to study the cosmic variance of the MW formation. In a first test, we require that the ensemble of 30 simulated merger trees should match with observations, i.e., we average over all 30 realisations. In a second step, we  ask if we can also constrain the SV, if we take the cosmic variance into account.

These merger trees are agnostic to the local value of the SV since they come from a DM-only simulation. Only the baryonic physics that we model with \asloth\ on top of these merger trees is affected by the SV.

\section{Results}
\subsection{Delay of star formation}
\begin{figure}
%/home/thartwig/asloth/scripts/utilities/plot_tgas_VBC.py
 \includegraphics[width=\columnwidth]{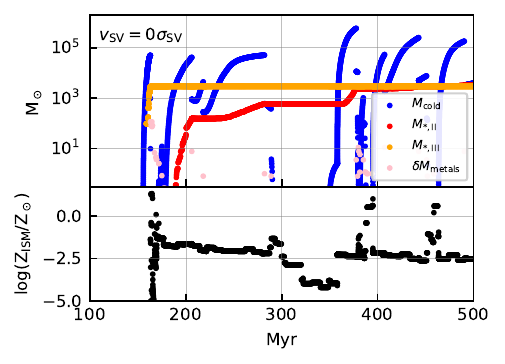}
 \includegraphics[width=\columnwidth]{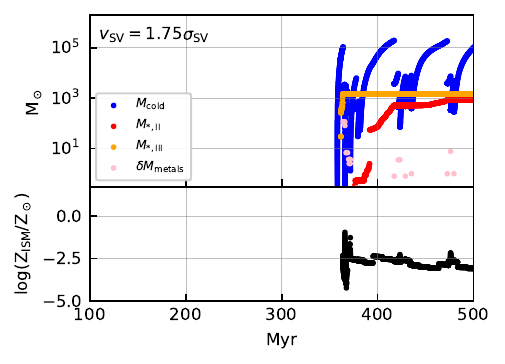}
 \caption{Onset of star formation in one merger tree branch for two different streaming velocities. Time is measured since the Big Bang, i.e., the shown interval corresponds to roughly $z=29$ to $z=10$. We see a clear delay of star formation of about $200$\,Myr for the higher SV (lower panel) compared to the run without SV (top panel). Pink points indicate the amount of metals that are ejected by SNe in this timestep in this halo.}
 \label{fig:tgas}
\end{figure}

We first show the effect of the SV for one halo as example in Fig.~\ref{fig:tgas}.
Orange and red points show the cumulative stellar mass of Pop~III and Pop~II stars that have formed in this specific branch. Blue points show the amount of cold dense gas inside a halo, which is strongly effected by SN feedback.
The onset of Pop~III star formation (orange) is delayed by about $200$\,Myr in the case of the higher SV. This is because the halo has to grow to a higher mass before it surpasses $M_\mathrm{crit}$ to trigger star formation.

We also see differences in the ISM metallicity when the second-generation stars form (red). The reason is that once the halo with the higher $v_\mathrm{SV}$ forms stars, the gravitational potential is deeper, and the halo can reaccrete gas more efficiently after the first SN explosions. This results in slightly lower ISM metallicities at the onset of second-generation star formation. While this is an interesting connection between the SV and metallicity of second-generation stars, we emphasize that this is just one halo in one merger tree and these quantitative conclusions might not always be valid.

The ISM metallicity fluctuates and sometimes even reaches super-solar values. These are periods in which SN feedback has ejected effectively all gas out of the halo and there is hardly any ISM left with which the freshly produced metals can mix. Therefore, the ISM metallicity appears very high in these episodes. However, the gas is not Jeans unstable in such situations and can not form stars. Only after the accretion of new IGM, the metallicity goes back to normal, the gas becomes Jeans unstable, and we can form stars again. So this ISM metallicity spikes are a numerical artefact and have no influence on the metallicity of stars.

\begin{figure}
 \includegraphics[width=\columnwidth]{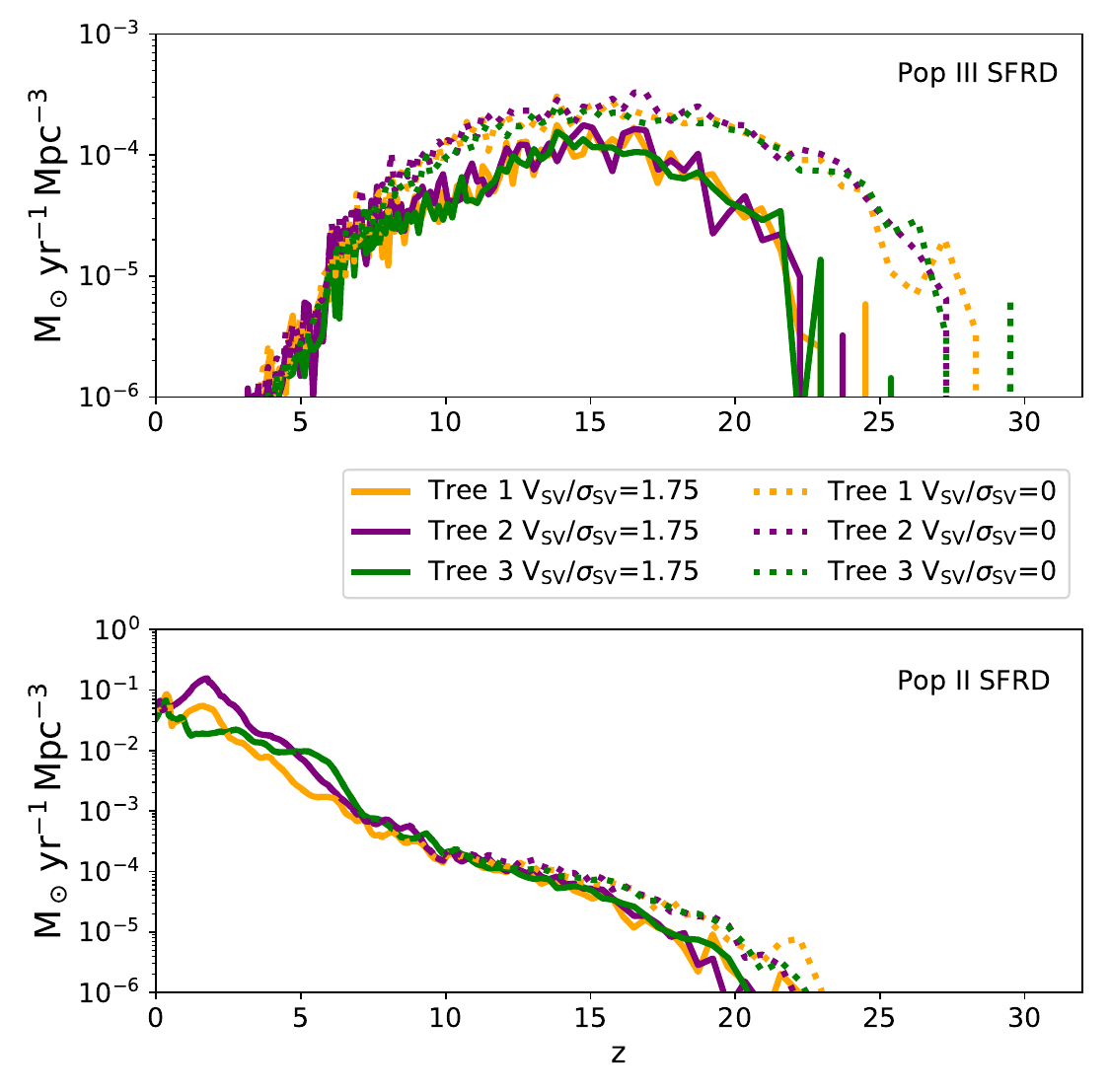}
 \caption{Star formation rate for Pop~III (top panel) and Pop~II stars (bottom panel). We simulated three different merger trees for two different SV. The blue, orange and green lines represent the SV $v_\mathrm{SV} = 1.75 \sigma_\mathrm{SV}$;
red, purple and brown lines for $v_\mathrm{SV} = 0\sigma_\mathrm{SV}$. We see a delay of star formation for both Pop~III and Pop~II star formation.}
 \label{fig:SFR}
\end{figure}

To study the effect of the SV on star formation in general, we show the SFRD of several merger trees in Fig.~\ref{fig:SFR}.
The volume here is considered to be the virial volume of the MW and we choose three arbitrary merger trees to illustrate that the effect of a SV is larger than random cosmic variance between merger trees. We see again a delay in star formation for the higher SV of on average $\sim 50$\,Myr, which is consistent with previous results \citep{greif11,OLeary12,richardson13,asaba16,schauer22}. This delay is due to the increased value of $M_\mathrm{crit}$.

We also notice that the SV only causes a systematic difference at $z>5$, whereas cosmic variance seems to dominate the differences between merger trees at lower redshift. Since the influence of the SV dominates at high redshift, it is crucial to consider observables that allow us to probe the high-z Universe.

\subsection{Optimal value of SV}
\begin{figure}
 \includegraphics[width=\columnwidth]{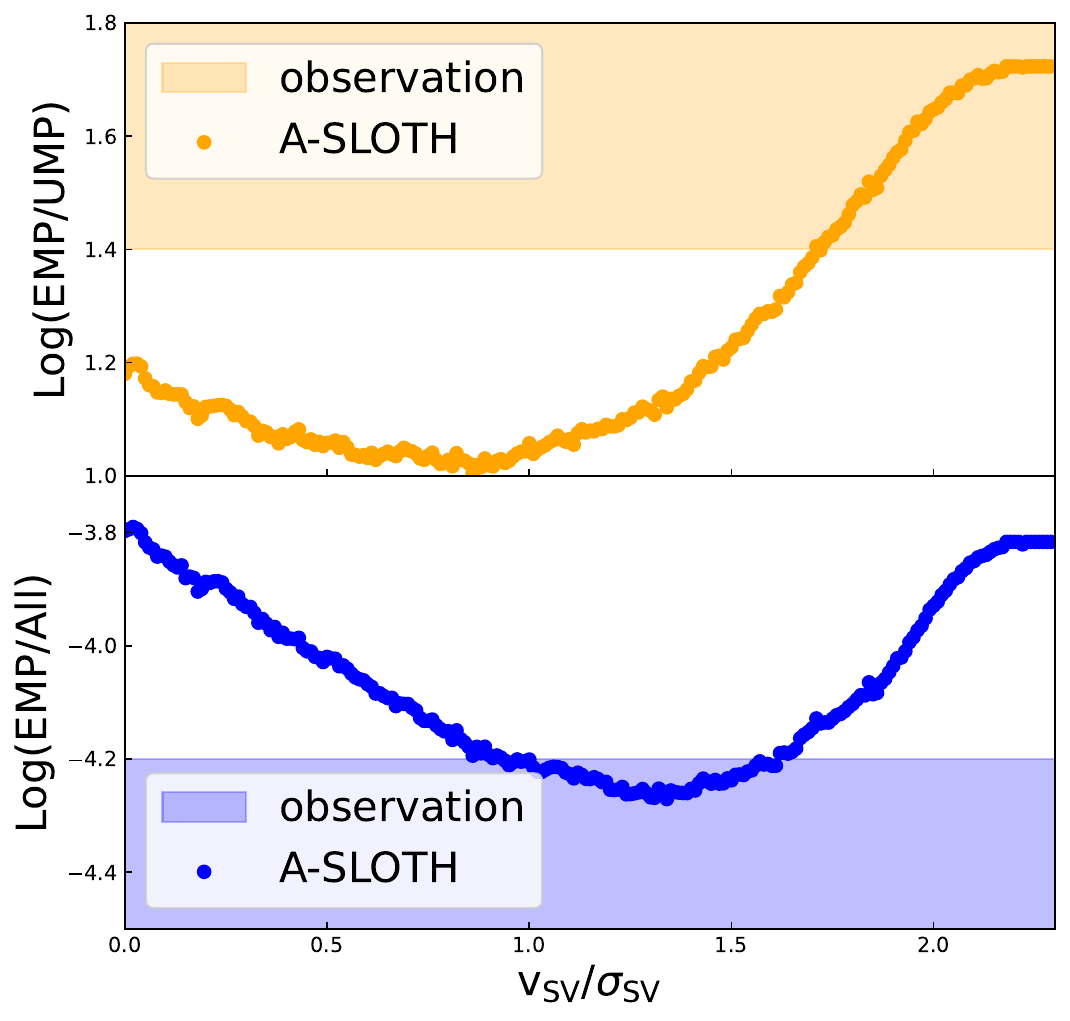}%A-SLOTH/Betul/why_emp/EMPUMP_plots.ipynb
 \caption{We show the ratio of EMP to UMP stars (top panel) and the ratio of EMP to all stars (bottom panel) as a function of the SV. The observed ranges of these quantities are indicated by the filled regions (range of 3 independent observations \citep{youakim20,bonifacio21,chiti21} for EMP/UMP and $1\sigma$ interval for EMP/All \citep{bullock05,bell08}, see Tab.~\ref{tab:BestFitobs}).}
 \label{fig:EMP2UMP}
\end{figure}

We explore different scenarios for the SV by sampling values between $v_\mathrm{SV} = 0\sigma_\mathrm{SV}$ and $v_\mathrm{SV} = 2.3\sigma_\mathrm{SV}$ in increments of $0.01\sigma_\mathrm{SV}$. As we show below, the EMP-based observables are most sensitive to the SV. Therefore, we first show how these depend on the SV in Fig.~\ref{fig:EMP2UMP}.
While the ratio of EMP to UMP stars matches best with observations for high SV, the agreement between observations and simulations is best for intermediate values for the ratio of EMP to all stars. We confirm that the total mass of UMP stars is a decreasing function with SV, and the mass of EMP stars has a local minimum at around $v_\mathrm{SV}$ = 1.3$\sigma_\mathrm{SV}$, and the total mass of stars is roughly constant with the SV. The SV and metallicity of stars are connected via the different delay times, which result in different metallicities in the moment of second-generation star formation (see e.g. Fig.~\ref{fig:tgas}).

\begin{figure}
 \includegraphics[width=\columnwidth]{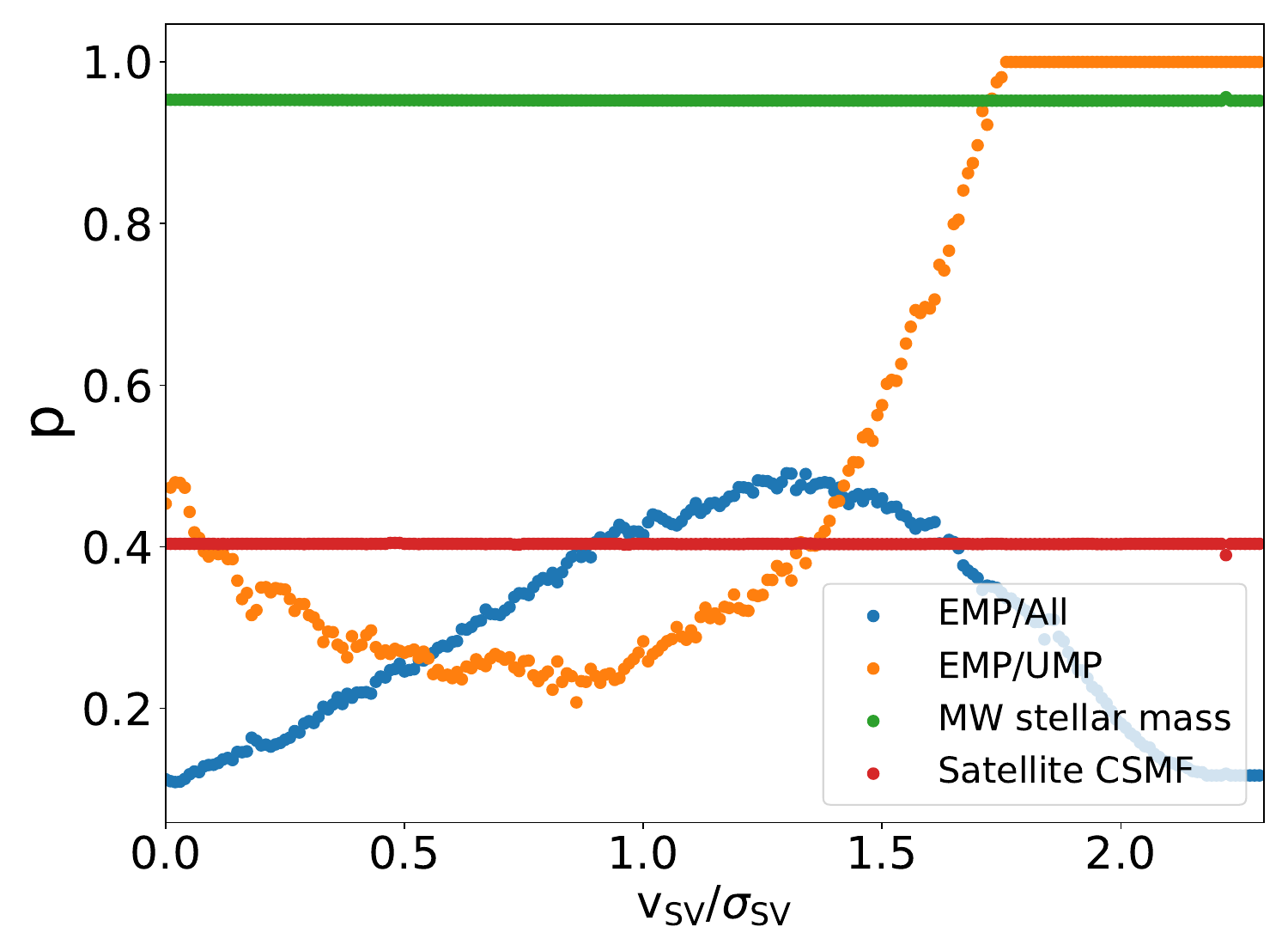}
 \includegraphics[width=\columnwidth]{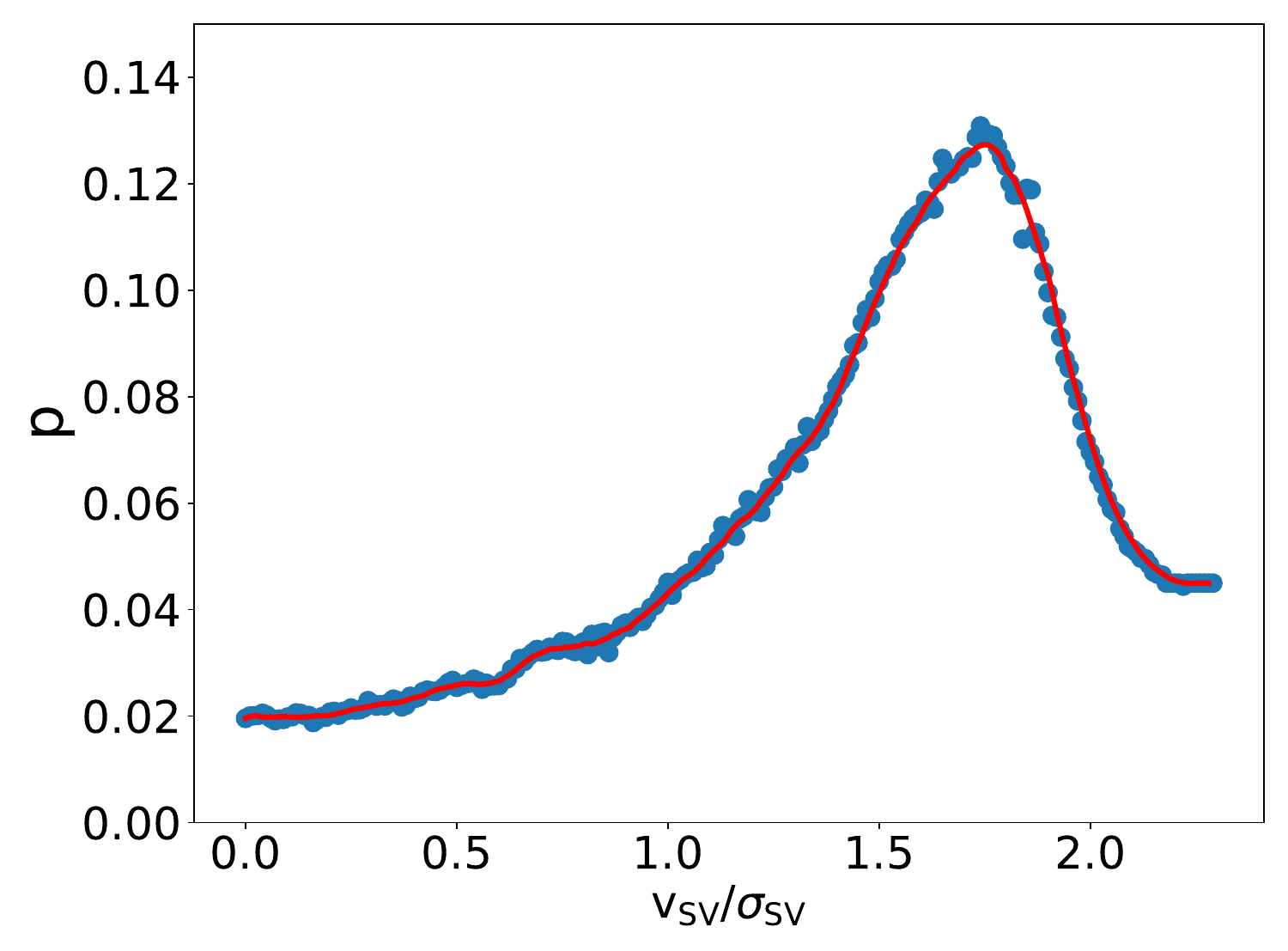}
 \caption{Top panel: p-values for different streaming velocity for four observables. The parameter p has the best fit at 1 and the worse fit at 0. The blue points represent the p for the ratio EMP to all stars, orange points: the fraction EMP to UMP in the MW, green points:stellar mass, red points: the CSMF. Bottom panel: Goodness-of-fit parameter as a function of the SV. The points represent the mean value of 30 \textit{Caterpillar} trees and the 4 observables. The red line shows the smoothing of the blue points, from which we can read the optimum SV of $v_\mathrm{SV}$ = 1.75$\sigma_\mathrm{SV}$.
}
 \label{fig:pSV}
\end{figure}

In Fig.~\ref{fig:pSV}, we quantify the fit to our four observables by showing the four individual p-values and the one combined goodness-of-fit parameter as a function of these sampled SVs. In the top panel, we see that two of the observables are almost constant. Therefore, we can state that the SV does not affect the stellar mass of the MW nor the stellar mass distribution of MW satellites, in agreement with the results by \citet{schauer22}.

In contrast, we can clearly observe that EMP and UMP stars are affected by the SV. The ratio of EMP to all stars has a local optimum at around $v_\mathrm{SV}$ = 1.3$\sigma_\mathrm{SV}$, and the ratio of EMP to UMP stars has a local minimum around $v_\mathrm{SV}$ = 0.8$\sigma_\mathrm{SV}$ and then increases to $p=1$ for higher values of $v_\mathrm{SV}$. It remains constant because the value produced by \asloth\ is then within the range of observed values. The distribution of the combined goodness-of-fit parameter (bottom panel) is a balance between the EMP to all and the EMP to UMP ratios.

\medskip

The main goal of the bottom panel is to find the highest $p$ for the combination of the four observables. By multiplying the four individual p-values, a profile is drawn from which it is possible to determine the optimal value of the SV. In order to determine this local SV, we also smooth the profile. A discussion of this smoothing can be found in section \ref{Smoothing}. We find the optimum local SV at the peak to be 
\begin{equation}
v_\mathrm{SV}=1.75\sigma_\mathrm{SV}.
\end{equation}
This result is the first estimate of the local value of the SV and it is crucial for the accuracy of MW simulations.

We summarize the best fit model in Tab.~\ref{tab:BestFitobs}. All four observables are in excellent agreement with the model predictions and we find individual p-values of $\geq 0.34$. The p-value for the CSMF is derived based on the Kolmogorv-Smirnov test (i.e. as difference between two cumulative probability functions), and therefore there is no specific value to be compared between observation and simulation.
\begin{table}
\centering
 \caption{List of the four observables, their literature values (if available), the best matching values from our model, and the resulting p-values.}
 \label{tab:BestFitobs}
 \begin{tabular}{lccc}
  \hline
  Observable & Observed & Simulated & p-value\\
  \hline
  log10(EMP/All) & $-4.7 \pm 0.5$ & -4.13 & 0.34\\
  log10(EMP/UMP) & $(1.4-2)\pm 0.3$ & 1.42 & 0.98\\
  CSMF & -- & -- &0.40\\
   M$_{*, \mathrm{MW}}/\Msun$& $(5.43 \pm 0.57) \times 10^{10}$ & $5.54 \times 10^{10}$ & 0.95\\
  \hline
 \end{tabular}
\end{table}

\subsection{Calibration and Global Optimum}
\begin{figure}
    \centering
 \includegraphics[width=\columnwidth]{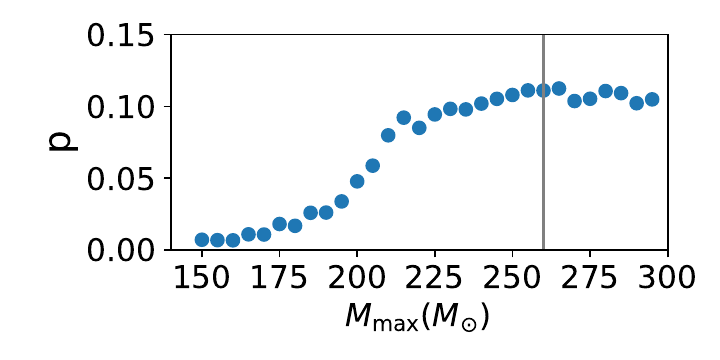}
 \includegraphics[width=\columnwidth]{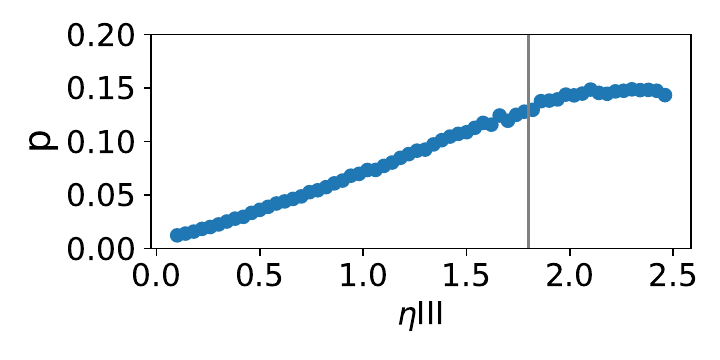}
 \includegraphics[width=\columnwidth]{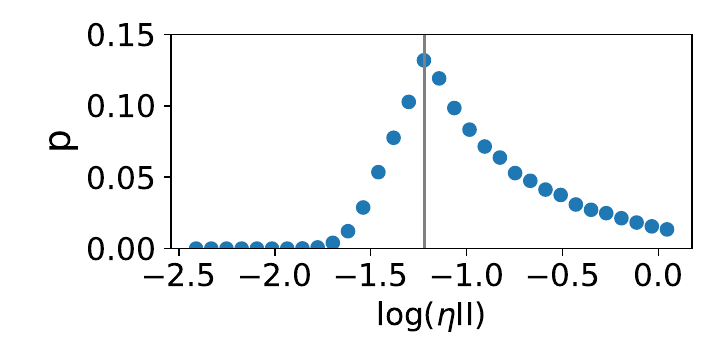}
 \includegraphics[width=\columnwidth]{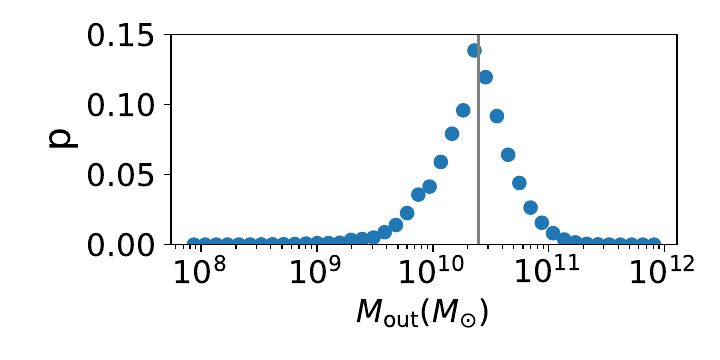}
 \includegraphics[width=\columnwidth]{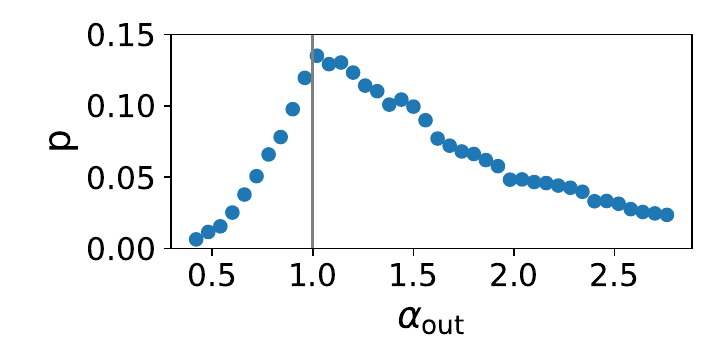}
 \caption{Fit quality as a function of one input parameter. For this explorations, we keep all other input parameters  fixed (including $v_\mathrm{SV}=1.75\sigma_\mathrm{SV}$) at their optimal values (vertical lines). This analysis demonstrates that we found a global optimum of our model calibration, and that the constrained value of $v_\mathrm{SV}$ is robust against changes in other input parameters.}
 \label{fig:5para}
\end{figure}

Our model has various input parameters that all need to be calibrated against observations. The goal of the calibration process is to find a global optimum, which guarantees that no other value of the SV will provide a better fit to the data. In Fig.~\ref{fig:5para} we therefore explore the five most important input parameters and show that our calibration found a global optimum.

In contrast to \citet{Hartwig22}, we do not vary the escape fraction of ionising photons, but keep them at their previously calibrated values of $f_\mathrm{esc,II}=0.37, f_\mathrm{esc,III}=0.6$. This is because that value was calibrated based on the optical depth to Thomson scattering in a cosmologically representative volume. The MW-like merger trees of this study are too small and therefore not able to reproduce this observable. It is therefore safest to use its previously calibrated value.

For all other parameters, we show their optimal values in Tab.~\ref{tab:BestFitpara}.
\begin{table}
\centering
 \caption{Best fit input parameters. Details and definitions of these can be found in \citet{Hartwig22}.}
 \label{tab:BestFitpara}
 \begin{tabular}{lcc}
  \hline
  Parameter & Symbol & Optimal Value\\
  \hline
  Streaming Velocity & $v_\mathrm{SV}$ & $1.75\sigma_\mathrm{SV}$\\
  \hline
  Maximum Pop~III mass & $M_\mathrm{max}$ & $260\Msun$\\

  \hline
  Pop III formation efficiency & $\eta_\mathrm{III}$ & 1.8\\
  \hline
  Pop II formation efficiency & $\eta_\mathrm{II}$ & 0.06\\
  \hline
   Normalization mass of outflow efficiency & $M_\mathrm{out}$ & $2.5\times10^{10}\Msun$\\
  \hline
  Slope of outflow efficiency & $\alpha_\mathrm{out}$ & 1.0\\
  \hline
 \end{tabular}
\end{table}
For the maximum mass of the Pop~III IMF, the Pop~II SFE, the normalisation mass, and slope of the feedback model, we can see that these chosen values in our fiducial model are exactly at their global optimum position (grey vertical lines). While the optimum for $M_\mathrm{max}$ is rather broad, we can not find any better value than the current model. For the Pop~III SFE, one could argue that a slightly higher value might provide an even better fit. However, the difference is not big and the improvement for the fit quality only minor. Moreover, inferring the global optimum in six dimension (these five input parameters and the SV) is challenging and the current parameters are already very close to the global optimum; definitely good enough to infer the local value of the SV. We will explore all input parameters and their posterior distributions simultaneously in a more detailed follow-up study.

\section{Discussion}
We have used the semi-analytical model \asloth\ to estimate the local value of the baryonic streaming velocity. We achieved this by simulating 30 MW-like galaxies with varying values for the SV. We have shown that a value of $v_\mathrm{SV}=1.75\sigma_\mathrm{SV}$ agrees best with 4 independent observations and is also robust against variations in other input parameters.

A local value of $v_\mathrm{SV}=1.75\sigma_\mathrm{SV}$ means that $\sim 97\%$ of the Universe has a smaller baryonic streaming velocity than the MW. Hence, the MW has formed out of a cosmic volume with rather high baryonic streaming velocity. We can also conclude that the star formation threshold for MW progenitor halos was close to the atomic cooling limit (compare Fig.~\ref{fig:halo_mass}). Therefore, only some pristine halos have formed stars via the molecular cooling channel in minihalos in our cosmic neighbourhood.

With a present-day variance of $\sigma_\mathrm{SV} (z=0) = 0.027\,\mathrm{km\,s}^{-1}$ we find that the absolute value of the SV in our current MW is around $v_\mathrm{SV} \approx 0.05\,\mathrm{km\,s}^{-1}$. This velocity is smaller than most characteristic velocities in astronomy, which emphasizes that the SV is only relevant at high redshift.

\subsection{Missing satellite problem}
\label{sec:MissingSat}
The dwarf galaxy problem (or missing satellites problem) is the conflict between the predicted number of satellite galaxies by the standard model of cosmology (cold DM) to the observed number. According to simulations, there are supposed to be more dwarf galaxies orbiting the MW than found by observations \citep{Simon_2007}. However, \citet{sales22} recently argued that this tension between observation and theory is not a problem anymore if one correctly accounts for baryonic physics and observational incompleteness.

\begin{figure}
\centering
 \includegraphics[width=\columnwidth]{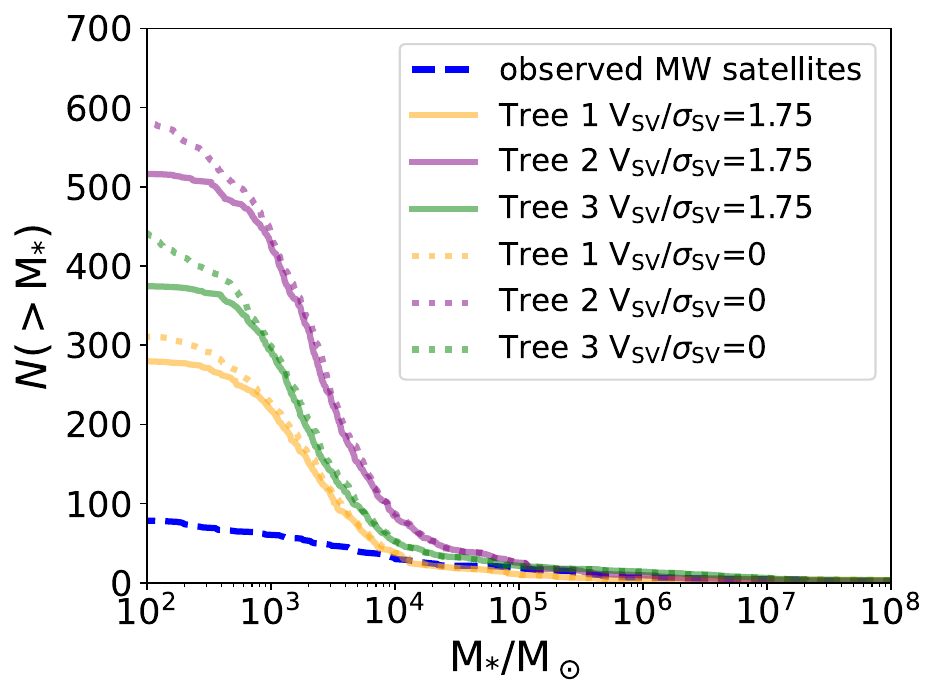}
 \caption{Cumulative number of MW satellite galaxies as a function of the stellar mass (CSMF). The three colors show three different merger trees. The solid line merger trees have been simulated for   $v_\mathrm{SV}=1.75\sigma_\mathrm{SV}$; and the pointed lines for V$_\mathrm{SV}$=0$\sigma_\mathrm{SV}$. The dark blue dashed line represent the observed CSMF, which is known as being incomplete \citep{McConnachie12,munoz18}.}
 \label{fig:CumulHalo}
\end{figure}

As the SV could be a potential solution to this tension \citep{maio11}, we compare the simulated number of satellite galaxies and the observed ones with and without the SV to evaluate its influence. In the Fig. \ref{fig:CumulHalo}, we plot the cumulative number of MW which we simulated for three different merger trees and for two different SV, namely $v_\mathrm{SV} = 1.75\sigma_\mathrm{SV}$ and $v_\mathrm{SV} = 0\sigma_\mathrm{SV}$. From this plot, we notice that the difference between two SV is smaller than the difference between two merger trees.
This means that the SV has not a big impact on the dwarf galaxy problem, and the number of MW satellite galaxies is more subjective to merger history rather than SV. In conclusion, the SV can mitigate the problem but not solve it. 

\subsection{Smoothing scale}\label{Smoothing}
To smooth the goodness-of-fit parameter in Fig.\ref{fig:pSV}, we use a convolution function. This latter convolves two arrays. In our case, we smooth over 10 neighboring points with a rectangular kernel. This choice has been made due to different test, showing that a smaller number of neighboring points gives a blurred smoothing, which makes it impossible to determine the optimum SV, while a larger number smooths it too much and gives a range of optimum SV.

Moreover, the effective volume out of which the MW has formed (2-6\,Mpc) might be slightly larger than the coherence length of the SV ($\sim 3$\,Mpc). This would mean that we can not assume so find just one constant value of the baryonic streaming velocity in each MW merger tree. To mimic the spread of SVs, we also use this smoothing kernel on a scale of $0.1 \sigma_\mathrm{SV}$ to erase small-scale fluctuations.

\subsection{Cosmic variance}
So far, we have taken all 30 \textit{Caterpillar} trees together and required that \asloth\ should reproduce the four observables for the ensemble of merger trees. We also took the tree-to-tree scatter as additional term of the variance when we calculated the goodness-of-fit parameters. However, the real MW is just one realisation of the DM merger history. So while our previous approach was statistically more robust, we also want to ask if we can infer the streaming velocity if we require that just one individual merger tree should reproduce the four observables.

\begin{figure}
    \centering
    \includegraphics[width=\columnwidth]{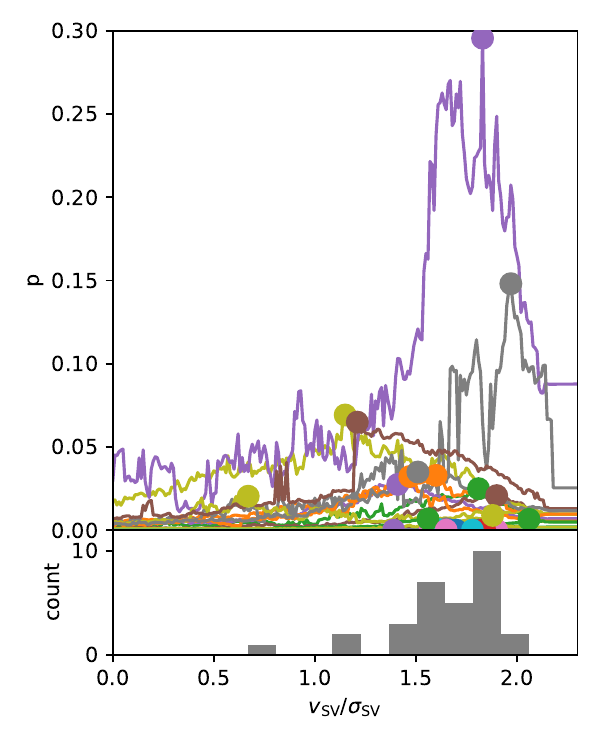}
    %/home/thartwig/Documents/A-Sloth/betul/tree2tree/tree2tree.ipynb
    \caption{Goodness-of-fit parameter as a function of the SV. The 30 \textit{Caterpillar} trees are plotted with different colors (same color may be used for up to 3 different trees). For each tree, we highlight the position of the maximum with a point and show the histogram of the positions of these points in the bottom panel. While some trees provide higher $p$ than other trees, their optimum is in a narrow range; specifically, the central 68\% of optimal values for the SV are in the range $1.47-1.88$.}
    \label{fig:indivi_trees}
\end{figure}

For this purpose, we repeat the experiment from before and calculated the goodness-of-fit parameter as a function of the SV. However, we now calculate $p$ separately for each of the 30 merger trees and show the results in Fig.~\ref{fig:indivi_trees}.
In the top panel, each line color represents one merger tree. On average, the goodness-of-fit parameters are smaller than before, because previously we added the cosmic variance as additional scatter term. We also highlight the position of the maximum of each merger tree by a point. These points mark the optimum for the SV of each merger tree and we show a histogram of these optimal SVs in the bottom panel.

The comic variance causes different DM merger histories, which all lead to similar present-day MWs. However, when we analyse the optimal SV for each tree, we still find that these values cluster around our previous optimum. Specifically, we find that the central 68\% of the data lies in the interval $1.47 \leq v_\mathrm{SV}/\sigma_\mathrm{SV} \leq 1.88$, and the central 95\% lies in the interval $1.15 \leq v_\mathrm{SV}/\sigma_\mathrm{SV} \leq 1.97$. So when we take into account the uncertainty due to cosmic variance, we can constrain the optimal value of the local SV to be
\begin{equation}
    v_\mathrm{SV} =1.75 ^{+0.13} _{-0.28}\,\sigma_\mathrm{SV}.
\end{equation}
Please note that these are central percentiles and do not assume that the distribution is Gaussian.

Out of 30 independent merger trees, only one requires an optimal value of $v_\mathrm{SV} =0.67\sigma_\mathrm{SV}$, and all 29 other trees require $v_\mathrm{SV} >1.1\sigma_\mathrm{SV}$. This means that ignoring the baryonic streaming velocity for simulations of the MW formation (i.e., $v_\mathrm{SV} =0\sigma_\mathrm{SV}$) is never ideal, and using the cosmologically most likely value ($v_\mathrm{SV} =0.8\sigma_\mathrm{SV}$), will also very likely underestimate the effect of the SV in the MW.

\subsection{Caveats}
We have shown that our results are robust with respect to the main assumptions, such as cosmic variance or the choice of other model input parameters. However, there are further effects that might influence the results and that we did not explicitly account for in this study.

The critical mass for star formation depends on the SV and on the LW background. Our simulated volume is too small to calculate the LW background self-consistently. Moreover, the strength of the LW background at a given redshift is a distribution with local variations \citep{hartwig19}. So while our choice of using an external LW background based on an analytical function is reasonable, a more detailed treatment of the LW background radiation \citep{trenti09,ahn09,LW23} could affect the critical halo mass and hence the optimal value of the SV.

To estimate the influence of the LW background, we can calculate the variations in $M_\mathrm{crit}$ for a change in the LW background in our fiducial model with $v_\mathrm{SV} =1.75\sigma_\mathrm{SV}$. At $z \leq 10$, the critical mass is dominated by the atomic cooling mass, which is independent of the LW background. At $z=20$, where star formation in minihalos can occur, a change of one order of magnitude in the LW background, results in a change of $M_\mathrm{crit}$ of only $20-30\%$. This shows that $M_\mathrm{crit}$ (and hence our results) are quite stable with respect to variations in the LW background.

We include the effect of the SV by increasing the critical halo mass for collapse based on the results by \citet{schauer_influence_2021}. In addition, the SV can also lower the gas mass fraction inside the first halos \citep{naoz_simulations_2013,richardson13,park20}, which we do not take into account in our model. However, this effect is only relevant in halos with masses $\lesssim 10^6\Msun$. All star-forming halos in our models are $> 10^6\Msun$. Hence, the assumption to initialise them with the cosmic mean baryon fraction is justified. One could also imagine that the SV influences the transport of metals in the IGM and might therefore effect the efficiency of external enrichment in the early Universe. However, this effect has not been studied nor quantified.

To model the dependence of $M_\mathrm{crit}$ on the SV, we use the fit to simulation results by \citet{schauer_influence_2021}. While this simulation is unprecedented in terms of numerical resolution and included physics, there are also other models. For example, \citet{kulkarni21} use a similar approach and find slightly different results for $M_\mathrm{crit}$ as a function of redshift, SV, and LW background. Our fiducial model finds $v_\mathrm{SV} =1.75\sigma_\mathrm{SV}$, which means that $M_\mathrm{crit}$ is close to the atomic cooling mass, at which point it becomes independent of the SV. Therefore, if we would repeat our analysis with a different model for $M_\mathrm{crit}$, we expect the optimal curve to be close to our fiducial model (blue dashed line in Fig.~\ref{fig:halo_mass}), it might just translate into a slightly different numerical value for the optimal SV.

We only explore one input parameter at a time, while we keep the other input parameters fixed. While this confirms that our optimal value of the SV is a local optimum, it can not guarantee that this is also a global optimum. Moreover, this method does not allow to investigate degeneracies between input parameters. Such degeneracies could widen the possible parameter range of all input parameters. We will investigate possible degeneracies and derive the full posterior distributions of all input parameters in a future study.

While these caveats might affect the quantitative conclusion, they will not alter our qualitative findings: the formation of EMP and UMP stars is influenced by the local SV and their observed numbers can hence be used to estimate the local value of the SV out of which the MW has formed. Further updates and improvement of our pilot study will help to strengthen also the quantitative conclusion in the future.

\subsection{Summary and Conclusion}
In this paper, we have studied the local SV and its effects on the formation of the MW. We found the local value of the SV to be $v_\mathrm{SV} =1.75\sigma_\mathrm{SV}$, which best reproduces observations. If we take into account the stochastic formation history of the MW, we find the optimal value to be in the range $1.47 \leq v_\mathrm{SV}/\sigma_\mathrm{SV} \leq 1.88$.

Our analysis confirms previous results that the SV delays the onset of star formation, but has no significant influence on the total stellar mass at redshift $z=0$. Neither in the satellite galaxies, nor in the main MW Galaxy. However, we find that the SV has an influence on the number of EMP and UMP stars, which can in turn be used to constrain the local value of the SV.

We have also investigated the influence of the SV on the missing satellite problem. While a high local SV can lower the number of luminous MW satellites, it can not completely solve the missing satellite problem.

Previous studies mainly focused on the influence of the SV on high-z galaxy formation. Our new results show that the SV and its correct local value can also have an observational consequence on the formation of the MW and its population of metal-poor stars. Our results can help to improve future models of MW formation.

\section*{Acknowledgements}
We thank Yurina Nakazoto, Huynbae Park, Anna Schauer, and Naoki Yoshida for valuable discussions. We are grateful to the referee for constructive feedback that has helped us to improve the manuscript. We acknowledge funding from JSPS KAKENHI Grant Number 20K14464.

\section*{Data Availability}
The underlying data will be shared upon reasonable request to the authors. The source code of \asloth\ is available online\footnote{\url{https://gitlab.com/thartwig/asloth}}.



\bibliographystyle{mnras}
\bibliography{MyBib} % if your bibtex file is called example.bib

%\begin{thebibliography}{99}
%\end{thebibliography}

%%%%%%%%%%%%%%%%%%%%%%%%%%%%%%%%%%%%%%%%%%%%%%%%%%

%%%%%%%%%%%%%%%%% APPENDICES %%%%%%%%%%%%%%%%%%%%%

\appendix

%%%%%%%%%%%%%%%%%%%%%%%%%%%%%%%%%%%%%%%%%%%%%%%%%%

% Don't change these lines
\bsp	% typesetting comment
\label{lastpage}
\end{document}